\documentclass[twocolumn]{aastex61}

\newcommand\aastex{AAS\TeX}

\accepted{July 10, 2017}

\submitjournal{AAS Journals}

\shorttitle{\aastex\ Magnetic Inflation and Stellar Mass}
\shortauthors{Han et al.}

\begin{document}

\title{Magnetic Inflation and Stellar Mass I: Revised Parameters for the Component Stars of the {\it Kepler} Low-mass Eclipsing Binary T-Cyg1-12664}

\correspondingauthor{Eunkyu Han}
\email{eunkyuh@bu.edu}

\author[0000-0001-9797-0019]{Eunkyu Han}
\affil{Department of Astronomy \& Institute for Astrophysical Research, Boston University, 725 Commonwealth Avenue, Boston, MA 02215, USA}

\author[0000-0002-0638-8822]{Philip S. Muirhead}
\affil{Department of Astronomy \& Institute for Astrophysical Research, Boston University, 725 Commonwealth Avenue, Boston, MA 02215, USA}

\author[0000-0002-9486-818X]{Jonathan J. Swift}
\affiliation{The Thacher School, 5025 Thacher Rd. Ojai, CA 93023, USA}

\author[0000-0002-1917-9157]{Christoph Baranec}
\affiliation{Institute for Astronomy, University of Hawai`i at M\={a}noa, Hilo, HI 96720-2700, USA}

\author[0000-0001-9380-6457]{Nicholas M. Law}
\affiliation{Department of Physics and Astronomy, University of North Carolina at Chapel Hill, Chapel Hill, NC 27599-3255, USA}

\author[0000-0002-0387-370X]{Reed Riddle}
\affiliation{California Institute of Technology, 1200 East California Boulevard, Pasadena, CA 91125, USA}

\author[0000-0001-8589-1938]{Dani Atkinson}
\affiliation{Institute for Astronomy, University of Hawai`i at M\={a}noa, Hilo, HI 96720-2700, USA}

\author{Gregory N. Mace}
\affiliation{McDonald Observatory \& The University of Texas, 2515 Speedway, Stop C1400, Austin, Texas 78712-1205 USA}

\author[0000-0002-0112-7690]{Daniel DeFelippis}
\affiliation{Department of Astronomy, Columbia University, 550 West 120th Street, New York, New York 10027}

\begin{abstract}

Several low-mass eclipsing binary stars show larger than expected radii for their measured mass, metallicity and age. One proposed mechanism for this radius inflation involves inhibited internal convection and starspots caused by strong magnetic fields. One particular eclipsing binary, T-Cyg1-12664, has proven confounding to this scenario.  \citet{Cakirli2013a} measured a radius for the secondary component that is twice as large as model predictions for stars with the same mass and age, but a primary mass that is consistent with predictions.  \citet[][]{Iglesias2017} independently measured the radii and masses of the component stars and found that the radius of the secondary is not in fact inflated with respect to models, but that the primary is, consistent with the inhibited convection scenario. However, in their mass determinations, \citet[][]{Iglesias2017} lacked independent radial velocity measurements for the secondary component due to the star's faintness at optical wavelengths. The secondary component is especially interesting as its purported mass is near the transition from partially-convective to a fully-convective interior.  In this article we independently determined the masses and radii of the component stars of T-Cyg1-12664 using archival {\it Kepler} data and radial velocity measurements of both component stars obtained with IGRINS on the Discovery Channel Telescope and NIRSPEC and HIRES on the  Keck Telescopes. We show that neither of the component stars is inflated with respect to models.  Our results are broadly consistent with modern stellar evolutionary models for main-sequence M dwarf stars and do not require inhibited convection by magnetic fields to account for the stellar radii.

\end{abstract}




\section{Introduction}\label{intro}

Double-lined spectroscopic eclipsing binary stars (SB2 EBs) enable accurate and precise measurements of stellar masses and radii. They provide critical tests of modern stellar evolutionary models as well as useful empirical relations between fundamental stellar properties, such as mass, radius, metallicity and age \citep[e.g.][]{Terrien2012, Kraus2015}. SB2 EBs that contain at least one low-mass main-sequence star ($M_{\star} \lesssim 0.7 M_{\sun}$) are especially useful for testing the treatment of convection and degeneracy in evolutionary models \citep[e.g.][]{Feiden2013}.  To date, several dozen low-mass SB2 EBs are known \citep[e.g.][]{Torres2002, Ribas2003, Bayless2006, Lopez-Morales2007, Vaccaro2007, Devor2008, Irwin2009, Morales2009a, Morales2009b, Rozyczka2009, Huelamo2009, Fernandez2009, Irwin2011, Kraus2011,  Birkby2012, Lee2013, Zhou2015}.  Many have larger radii than predicted by evolutionary models for their mass, effective temperature and age. \\ 
\indent A leading theory for the radius discrepancy involves effects from strong magnetic fields.  In this scenario, rapid rotation produces strong magnetic fields within the star via the dynamo mechanism.  The magnetic fields inhibit convection within the star and create starspots on the surface.  A result of inhibited convection and starspots is a larger main-sequence radius and effective temperature for a given initial mass and metallicity \citep[e.g.][]{Chabrier2007, MacDonald2013}. This effect would be preferentially seen in EBs because of observational biases: short-period binary stars are more likely to eclipse, so most eclipsing binaries have short orbital periods (P $<$ 5 days). With short orbital periods, they are calculated to be tidally locked with rapid rotation.  Studies show a strong correlation between rapid rotation and magnetic activity for single stars, implying that rapid rotators in EBs likely also have strong surface magnetic fields \citep[][]{West2015}.

In the context of models, the effect of magnetic fields on stellar radius depends largely on the mass of the star, with less-massive, fully-convective stars ($M_{\star} \lesssim 0.35 M_{\sun}$) less affected than higher-mass, partially convective stars \citep[][]{Feiden2013}. Therefore, empirically measuring magnetic inflation vs. stellar mass is extremely useful to these modeling efforts and may even present a method for empirically determining the mass corresponding to the partially-to-fully convective boundary. \\ 
\indent However, recent studies have shown that such strong magnetic fields are not feasible in the low-mass stars. By  using the the magnetic Dartmouth stellar evolution code to reproduce the observed properties of the fully-convective detached eclipsing binary stars CM Draconis and Kepler-16, \citet{FC2014} found that for a star to be inflated due to the magnetic fields, the strength of the field has to be greater than 50 MG in the stellar interior to sufficiently alter convection, which is subject to rapid decay due to magnetic buoyancy instability, macroscopic diffusion, and advection from the convective medium. 
\citet{Browning2016} independently found that for a 0.3 $M_{\sun}$ star, flux tubes with magnetic fields stronger than 800 kG are not sustainable in the stellar interior. Using collections of thin flux tubes and assuming a simple magnetic morphology, they investigated the timescale of the dissipation of strong magnetic fields due to the magnetic buoyancy instabilities and Ohmic dissipation. For the magnetic fields structured on small-scales, the regeneration of the fields are faster than the destruction by buoyancy instability whereas for the large-scale fields, field loss from the buoyancy instability are faster than the regeneration of the field. However, the small-scale magnetic fields are also susceptible to the Ohmic dissipation, which produces dissipative heat that is greater than the luminosity of the star. In both small and large-scale field configurations, strong magnetic fields are not feasible in the interiors of low-mass stars. \\
\indent Other proposed mechanisms for the radius discrepancy involve effects from stellar metallicity. In this scenario, metal-rich stars have a higher number density of molecules in their atmospheres, which keep heat within the star, ultimately increasing the radius of the star to conserve flux \citep[][]{Lopez-Morales2007}. CM Draconis was known to have a larger radius than model predictions, and empirical metallicity measurements of the component stars show that it actually is a metal-poor system ([Fe/H] = -0.3), providing evidence against this proposed scenario \citep[][]{Terrien2012}. However, a more recent study found that the metallicity of CM Draconis is near-solar, resulting an inflation of $\sim$2\% compared to the stellar evolutionary models \citep[][]{Feiden2014b}. \\
\indent Regardless of the predictions for the radii of low-mass stars, it is critically important that observers report accurate mass and radius determinations for low-mass SB2 EBs, as the measurements directly inform our understanding of the physical properties of stars in general and are used in relations that determine the physical properties of exoplanets found to orbit isolated stars. In this work, we revise the measured masses and radii for one such SB2 EB: T-Cyg1-12664,\footnote{$\alpha$=297.9159$^\circ$, $\delta$=+48.3321$^\circ$} or KIC 10935310. T-Cyg1-12664 was initially discovered by the Trans-Atlantic Exoplanet Survey \citep[TrES,][]{TrES}, a photometric survey for transiting exoplanets. \cite{Devor2008a} performed an automated search for EBs in TrES photometric data and found 773, one of which was T-Cyg1-12664. They classified T-Cyg1-12664 as an EB system with the orbital period of $\sim$8.2 days and two equal-sized component stars. Follow up spectroscopic observations revealed that T-Cyg1-12664 is not two equal mass stars but consists of a primary and secondary with a mass contrast of 1.9. \cite{Devor2008} revised the orbital period to 4.1 days and acquired six primary and one secondary radial velocity measurement; however, he did not report radii for the component stars.  \\
\indent \cite{Cakirli2013a} revisited and characterized T-Cyg1-12664 using {\it Kepler} data containing both primary and secondary eclipses and independently measured SB2 radial velocities. In their paper, \citet{Cakirli2013a} reported that the primary component of T-Cyg1-12664 has a mass and radius of 0.680 $\pm$ 0.021 $M_{\sun}$ and 0.613 $\pm$ 0.007 $R_{\sun}$ respectively, and that the secondary component has a mass and radius of 0.341 $\pm$ 0.012 $M_{\sun}$ and 0.897 $\pm$ 0.012 $R_{\sun}$, all with an age of 3.4 Gyr. If true, the secondary component would have a  significantly inflated radius compared to predictions for main-sequence stars of that mass and age, by well over a factor of two.  We note that T-Cyg1-12664 also appears in the {\it Kepler} Eclipsing Binary Catalog \citep[][]{Prsa2011, Slawson2011} but without stellar mass or radius determinations, and also in a catalog by \citet[][]{Eker2014}, but with masses and radii similar to \citet{Cakirli2013a}. \\
\indent One possible explanation for the large radius of the secondary is that it is a pre-main sequence star still undergoing contraction.  Kelvin-Helmholtz contraction timescales increase dramatically with lower mass, such that the primary may be on the main-sequence while the secondary remains pre-main-sequence, similar to the low-mass EB UScoCTIO 5 recently discovered by \citet{Kraus2015}. However, \citet{Cakirli2013a} estimated an age of the system to be 3.4 Gyr based on the characteristics of the primary star. After 1 Gyr, a 0.341 $M_{\star}$ star would have long settled onto the main-sequence.  Instead, the authors suggest the fully-convective, or near fully-convective, nature of the secondary star is related to the radius inflation. \\
\indent More recently, \cite{Iglesias2017} revisited T-Cyg1-12664. They independently analyzed the {\it Kepler} light curve, acquired their own optical photometric data (V, R, and I band) and independently measured SB1 radial velocities. Using the \texttt{PHOEBE} code \citep[][]{phoebe}, they revised the radii of both stars as well as the mass of the secondary star.  They found that the primary star is consistent with a G6 dwarf with a mass of 0.680 $\pm$ 0.045 $M_{\sun}$ and a radius of 0.799 $\pm$ 0.012 $R_{\sun}$ and that the secondary star is consistent with an M3 dwarf with a mass of 0.376 $\pm$ 0.017 $M_{\sun}$ and a radius of 0.3475 $\pm$ 0.0081 $R_{\sun}$. If true, the {\it primary} star would be inflated and the secondary would not be inflated with respect to magnetic-free evolutionary models.  Their results are broadly consistent with the magnetic inflation scenario, in which magnetic fields have a larger effect on the radii of higher-mass stars compared to lower-mass stars.  However, due to the faintness of the secondary star, \cite{Iglesias2017} were not able to measure radial velocities of the secondary star from their optical spectra, and thus their mass measurements rely on the radial velocity measurements of \cite{Cakirli2013a}.\\
\indent T-Cyg1-12664 could serve as a benchmark EB system if the mass of the secondary component is indeed 0.376 $M_{\sun}$, since the mass is near the transition from a partially to a fully convective stellar interior. As we show in the following sections, we independently determined the masses and the radii of each component of T-Cyg1-12664, and our measurements differ significantly from the previous two groups' measurements. We obtained independent SB2 radial velocity measurements, including infrared observations, and we re-analyzed the {\it Kepler} light curve.  We measure a mass of 0.92 $\pm$ 0.05 M$_{\Sun}$ and a radius of 0.92 $\pm$ 0.03 R$_{\Sun}$ for the primary star. For the secondary star, we measure a mass of 0.50 $\pm$ 0.03 $M_{\Sun}$ and a radius of 0.47 $\pm$ 0.04 $R_{\Sun}$. We attribute the difference in mass and radius determinations to our independent SB2 radial velocity measurements.  Our results are broadly consistent with modern stellar evolutionary models for main-sequence M dwarf stars and do not require inhibited convection by magnetic fields to account for the stellar radii. In \S \ref{Data} we describe the data used in our determinations. In \S \ref{model} we describe our modeling procedure and results.  In \S \ref{Discussion} we discuss the implications for the new mass and radius.

\section{Data}\label{Data}

\subsection{{\it Kepler} Light Curve}\label{lc}
\begin{figure}
\centering
\includegraphics[width=\columnwidth]{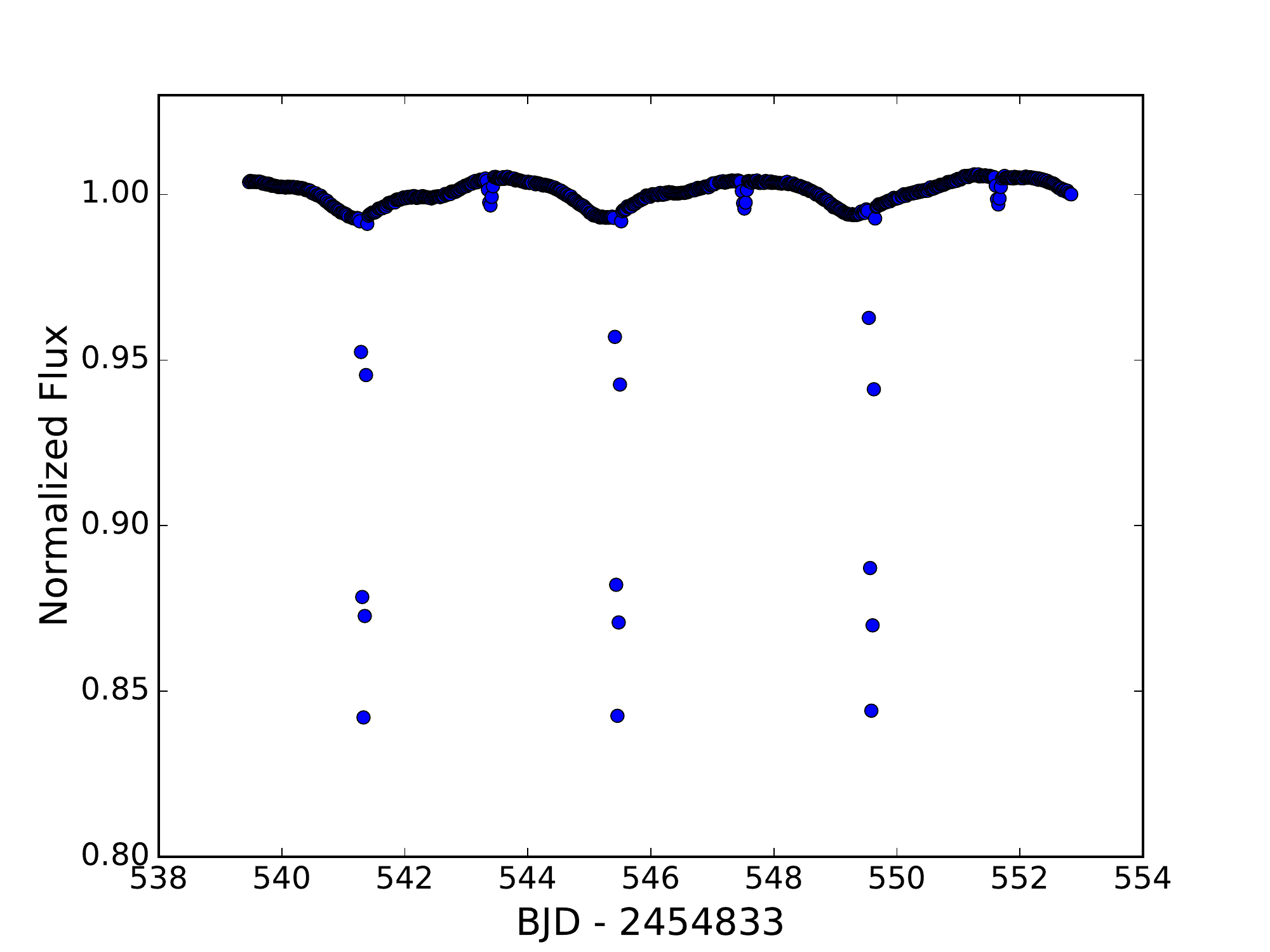}
\caption{Example of the {\it Kepler} long-cadence data showing three primary and secondary eclipses from quarter 6. Black points represent the out-of-eclipse flux and the blue points represent in-eclipse flux.  The out-of-eclipse modulation is consistent with star spots and spin-orbit synchronous rotation of either the primary or secondary component star.  The 30-min exposure times of {\it Kepler} long-cadence data provide only a half-dozen data points across each individual eclipse event.}
\label{kepler_sample}
\end{figure}

We obtained {\it Kepler} light curve data for quarters 1 through 17 from the Mikulski Archive for Space Telescopes (MAST). Long cadence data recorded at regular intervals and with exposure times of 1766 seconds are available for all quarters of the primary {\it Kepler} mission except for quarters 7, 11, and 15.
No short cadence data are available, which is contradictory to what is reported in \cite{Cakirli2013a}.  On inspection, the {\it Kepler} light curves show roughly 100 primary and secondary eclipses with a period consistent with the orbital period reported in \citet{Cakirli2013a}.  The light curves show out-of-eclipse modulation that is nearly synchronous with the system orbital period, reaching a maximum peak-to-peak amplitude of $\sim3\%$.  We attribute the modulation to starspots on the primary star combined with synchronous stellar rotation.  We used the PDCSAP\_FLUX data, which is corrected for effects from instrumental and spacecraft variation \citep[][]{Stumpe2012, Smith2012}. We also removed obvious outliers by hand. Figure \ref{kepler_sample} shows the first three primary and secondary eclipse pairs in quarter 6 of {\it Kepler} data. \\

\subsection{SB2 Radial Velocity Data}

\subsubsection{IGRINS Observations}

We observed T-Cyg1-12664 using the the Immersion GRating INfrared Spectrometer \citep[IGRINS,][]{Yuk2010} on the 4.3-meter Discovery Channel Telescope (DCT) on the nights of UT 2016 October 16 through UT 2016 October 18. IGRINS is a cross-dispersed, high-resolution near-infrared spectrograph with wavelength coverage from 1.45 to 2.5 $\mu m$. IGRINS has a spectral resolution of R = $\lambda/\Delta\lambda$ = 45,000 and allows for simultaneous observations of both H- and K-band in a single exposure \citep[][]{Yuk2010, Park2014, Mace2016}. The exposure times were calculated to achieve a signal-to-noise ratio of $\sim10$ or higher per wavelength bin. We observed A0V standard stars (HR 7098 and HD 228448) that are within 0.2 airmasses of T-Cyg1-12664, before or after target observations, for the purpose of telluric corrections. There is a publicly available reduction pipeline for the IGRINS \citep[][]{plp}, and we used this pipeline to process all of our spectra.  IGRINS is a visiting instrument at the DCT whose principal site is the McDonald Observatory in Texas. \\

\indent Cross-correlation templates with spectral types between G1 and M4 were also observed with IGRINS on the 2.7-meter Harlan J. Smith Telescope at McDonald Observatory, and reduced in the same manner as T-Cyg1-12664. Radial velocities for the template stars were determined using the method summarized in \cite{Mace2016} and are precise to 0.5 km/s. \\
\indent IGRINS' H- and K-band data contain 28 and 25 orders, respectively. The pipeline performs dark subtraction and flat-fielding first, followed by an AB subtraction to remove the OH airglow emission lines, and finally extracts the spectrum. The pipeline uses telluric airglow emission lines for wavelength calibration. However, the current pipeline version does not support careful removal of telluric absorption lines, so we further processed the pipeline extracted 1-D spectra. For this task, we used \texttt{xtellcor\_general}, a software tool designed to remove telluric lines from  near-infrared spectra \citep[][]{xtellcor}. The software accepts a measured spectrum of an A0V star and a target spectrum. It uses a model spectrum of Vega (an A0V star) to construct the telluric spectrum, calculates the relative shift between the observed A0V standard and the target spectrum, and applies the shift to the constructed telluric spectrum. In its final step, \texttt{xtellcor\_general} divides the telluric spectrum from the target spectrum. 

\begin{figure*}
\begin{center}
\includegraphics[width=0.75\linewidth]{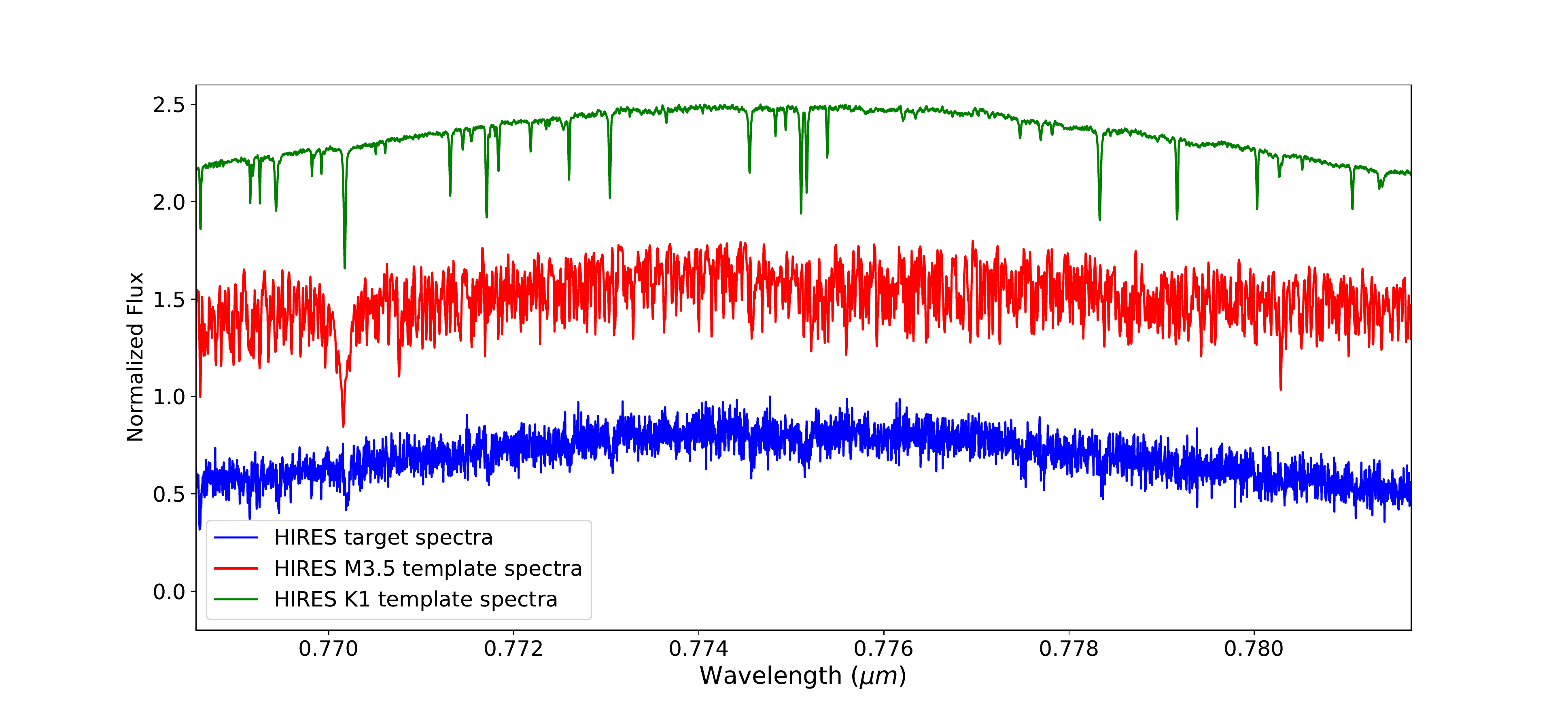}
\includegraphics[width=0.75\linewidth]{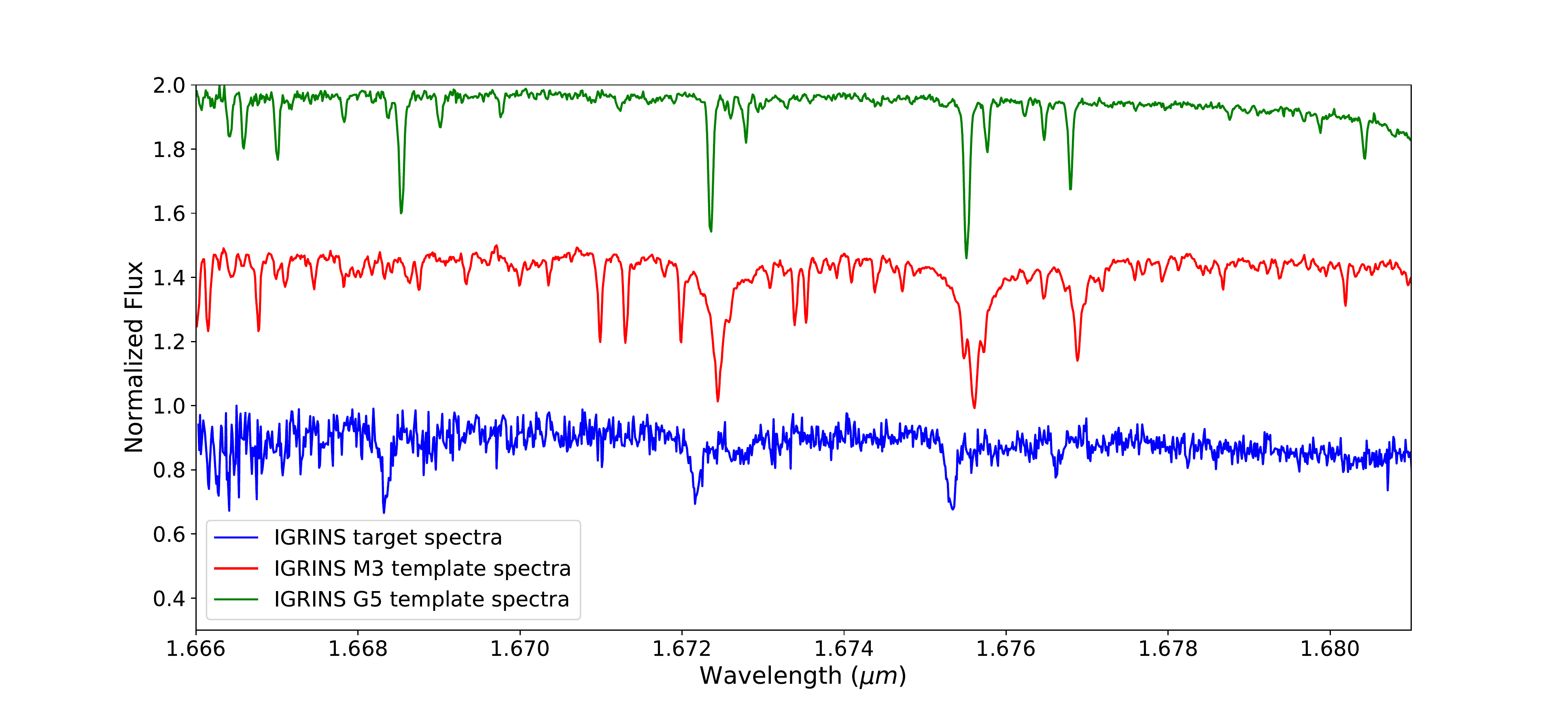}
\includegraphics[width=0.75\linewidth]{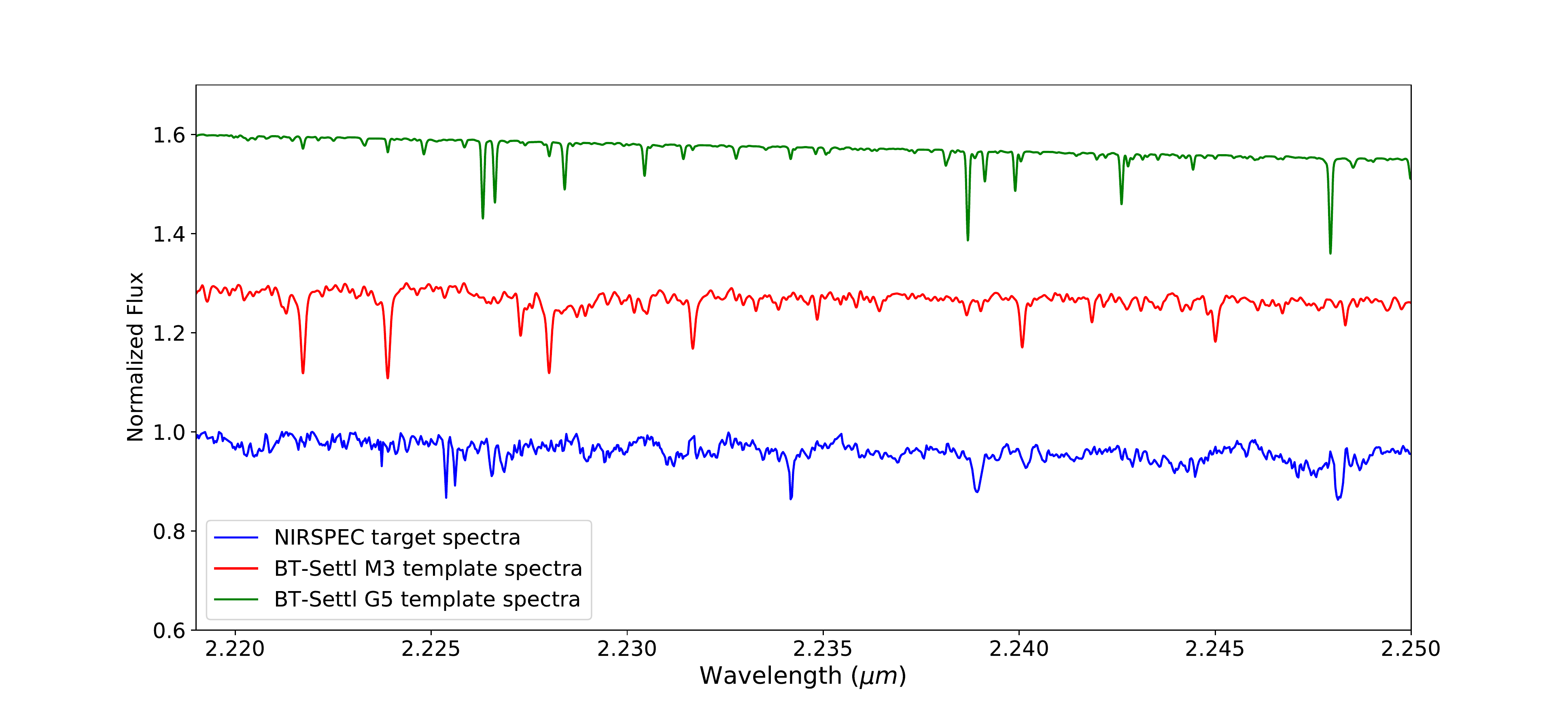}
\caption{Example spectra from HIRES (top), IGRINS (middle), and NIRSPEC (bottom). Radial velocity templates are plotted for comparison. Each plot shows a single order from the respective instrument.}
\label{sample_igrins}
\end{center}
\end{figure*}

The middle panel in Figure \ref{sample_igrins} shows a sample IGRINS H-band telluric-corrected spectrum. We processed all the target spectra and the radial velocity standard spectra to remove telluric lines. We selected the radial velocity standards in consideration of previously reported spectral types for each component star. \cite{Cakirli2013a} reported specral types of K5 and M3 and \cite{Iglesias2017} reported spectral types of G6 and M3. We used a G5 template as the radial velocity standard for the primary component (HIP 102574, with radial velocity of -69.8km/s) and an M3 template as a radial velocity standard for the secondary component (GJ 752, with radial velocity of 36.6km/s).

We only used IGRINS H-band data as the sky background in the K-band reduced the signal-to-noise of the reduced spectra. Of the 28 orders in the H-band spectra we selected 8$^{th}$ through the 14$^{th}$ due to their signal to noise. These orders gave us a wavelength coverage of 1.59 $\mu m$ to 1.70 $\mu m$.

\begin{figure}
\begin{center}
\includegraphics[width=\linewidth]{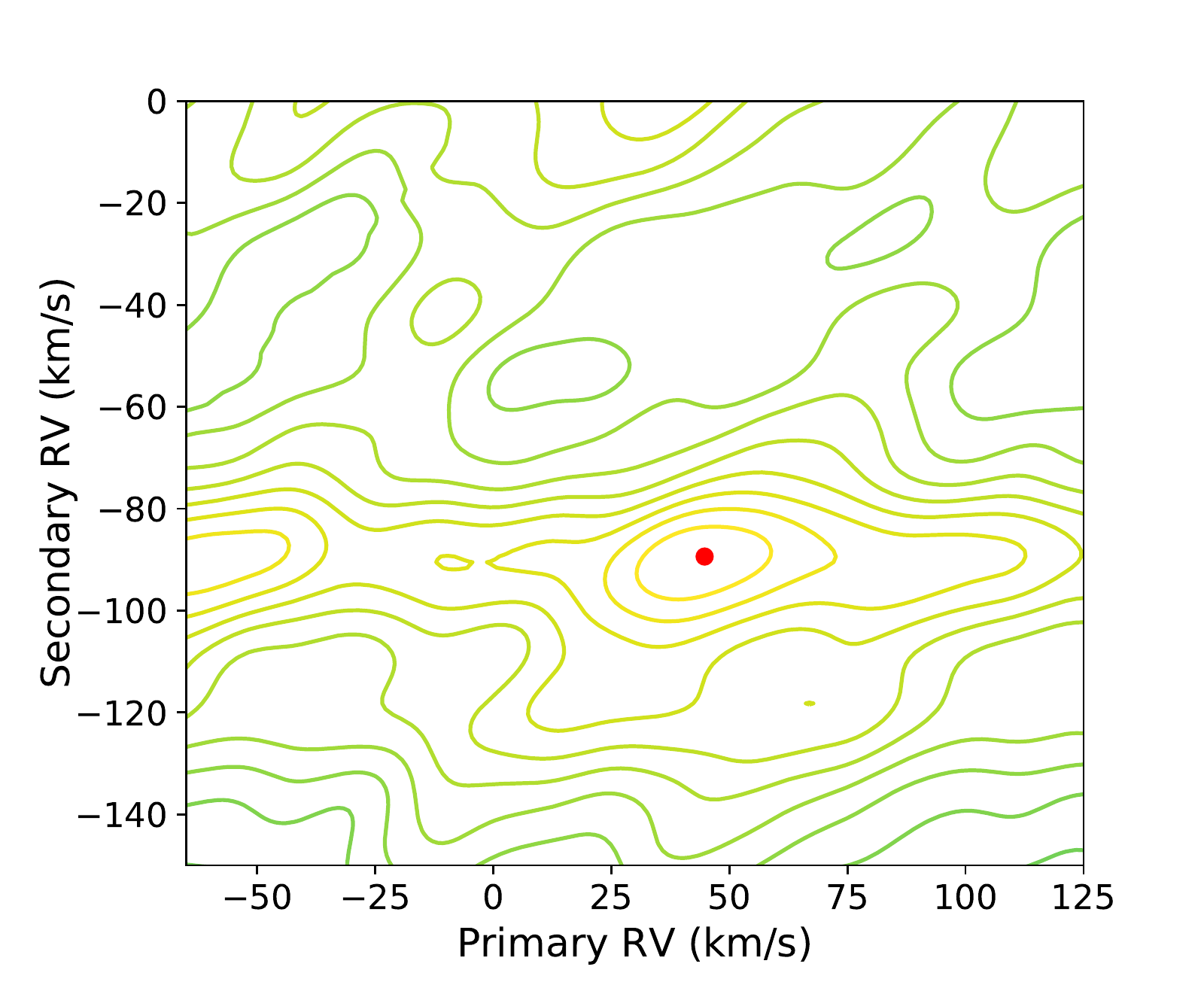}
\caption{A sample contour plot of the two-dimensional cross-correlation function using the NIRSPEC data. The red dot shows the location of the maximum value of the two-dimensional cross-correlation function. The corresponding primary and secondary axes are the calculated radial velocities of each component. }
\label{sample_todcor}
\end{center}
\end{figure}

To measure the radial velocity we cross-correlated the target spectra with the template spectra. Before cross-correlating, we transformed the wavelength scale of the spectra from regular intervals to logarithmically increasing intervals, so that radial velocity shifts in the spectra are equivalent to the same fractional shift in the wavelength interval regardless of the order. We interpolated the spectra onto the logarithmically increasing wavelength grid using a linear spline function. Next, we used the TwO-Dimensional CORrelation software package \citep[TODCOR][]{TODCOR} to measure radial velocities of each component star. TODCOR simultaneously calculates the radial velocities of each component by cross-correlating the target spectrum against the two templates over a range of radial velocities. This produces a two-dimensional cross-correlation function, with the peak location corresponding each component's measured radial velocity. We ran TODCOR on each order, using the mean between the orders as our measured radial velocities.  We estimated the uncertainties by calculating the root-mean-square of the radial velocities across the orders, and divided by the square root of number of orders used. Figure \ref{sample_todcor} shows an example two-dimensional cross-correlation function.  We calculated the barycentric Julian date for each observation, converted the radial velocities into the reference frame of the solar system barycenter for both the target and the radial velocity template, and report those as the final radial velocity measurements.

\subsubsection{NIRSPEC Observations}

We observed T-Cyg1-12664 with NIRSPEC on the W. M. Keck II Telescope \citep[][]{McLean1998} on the nights of UT 2014 July 06 and 2014 July 13. The first night was mostly cloudy with the average seeing of 0\farcs{}5 and the second night had some cirrus clouds with stable seeing between 0\farcs{}3 and 0\farcs{}5. NIRSPEC is a cross-dispersed near-infrared spectrograph with wavelength coverage from 0.95 to 5.5 $\mu m$. We used the high-resolution mode with a spectral resolution of R = $\lambda/\Delta\lambda$ = 25,000 and observed in the K-band with an ABBA nodding pattern. We observed A0V standard stars on each night (HD 203856 and HR 5984, respectively) that are within 0.2 airmasses of T-Cyg1-12664, before the target observations, for the purpose of telluric corrections.

To reduce the data, we used \texttt{REDSPEC}, a publicly available IDL based reduction pipeline for NIRSPEC \citep[][]{REDSPEC}. \texttt{REDSPEC} processes dark subtraction, flat fielding and rectification on each A and B frame, performs the AB subtraction, and extracts the 1D spectrum. \texttt{REDSPEC} uses Th, Ne, Xe, and Kr arc lamps to calculate the wavelength solution. However, for some orders, the arm lamp lines did not give precise wavelength solution due there being less than 3 prominent lines present in the order. Therefore, after reducing the spectra using \texttt{REDSPEC}, we used a custom script to correct the wavelength solution. We compared telluric absorption lines in A0V spectra acquired each night to the ATRAN model of telluric lines \citep[][]{ATRAN}, and calculated shifting and a stretching parameters to apply to the wavelength solution by minimizing $\chi^2$. Then we applied the two parameters to correct the wavelength solution of each target spectrum. After the wavelength corrections, we used \texttt{xtellcor\_general} and performed the same procedure as we did for the IGRINS data. We used BT-Settl models \citep[][]{Allard2012} corresponding to G5 and M3 spectral types as radial velocity templates as we did not have template spectra observed with NIRSPEC. We matched the spectral resolution of the model to the NIRSPEC data but did not apply the rotational broadening. We also found an additional NIRSPEC observation from the night of 2007 July 30 on the Keck Observatory Archive (KOA)\footnote{https://koa.ipac.caltech.edu/cgi-bin/KOA/nph-KOAlogin} and have included it in our analysis. To calculate the radial velocity, we performed the same method as we did with IGRINS data.

\subsubsection{HIRES Observations}
We obtained spectra using the HIRES echelle spectrometer on the W. M. Keck I telescope between UT June 11 and July 27, 2014 in partnership with the California Planet Search (CPS) program. The spectra had low signal-to-noise—between 2 and 5 per pixel—to minimize integration times and maximize phase coverage for the amount of time available. The C2 decker was used providing a 14'' x 0.861'' slit, when projected on the sky, translating to a spectral resolution of R $\sim$ 45,000. Integration times varied between 28 and 123 seconds to obtain approximately 1000 counts in the HIRES exposure meter. In some cases a maximum exposure time of 60 seconds was enforced, regardless of the exposure meter.

We followed the reduction process of \citet[][]{Chubak2012}, but made small adaptations to the code to accommodate the lower signal-to-noise observations. After finding a wavelength scale from Thorium-Argon calibration spectra, we mapped the spectra onto a logarithmic wavelength scale so that pixel shifts correspond to uniform shifts in velocity. We then used the telluric A and B molecular oxygen absorption bands to determine a wavelength zero-point for each spectrum by comparison to B star calibration spectra taken at the beginning of each night.

Of the 10 orders available on the red chip of HIRES, two were ignored due to profuse telluric absorption, and only half the 3rd and 8th orders were considered due to the A and B telluric bands. The first and last 50 channels of each band were also ignored to mitigate edge effects from the continuum fitting.  We fitted a continuum using a 3rd order polynomial, with the spectra binned by a factor of 10 to flatten.

Doppler measurements were carried out independently in each of the 8 remaining orders by minimizing chi-squared as a function of Doppler shift between the spectra and a template reference star taken from the \citet[][]{Chubak2012} program. To obtain radial velocities for the primary component a high SNR spectrum of the K1 dwarf HD 125455 (with radial velocity of -9.86km/s) was used. For the secondary component we used a template spectrum of the M3.5 dwarf GL 273 (with radial velocity of 18.21km/s). We applied barycentric corrections to the velocity for each spectrum, and then inspected the chi-squared function which was sampled at shifts of 0.1 pixel. We compared the radial velocities measured by this method with those measured with TODCOR using the M3.5 dwarf GL 273 and the G2 HD 146233 templates. The results were consistent within the uncertainties.

In each order, a Gaussian was fit to the chi-squared function to estimate the true minimum. The final radial velocity reported is the average of the radial velocities found in each order and the error reported is the standard deviation of the values. For some spectra, there was only one order in which the Gaussian fit did not fail precluding an estimation of the RV error by our chosen method. For these spectra we took the largest measured error of the secondary radial velocity, which was 6.8 km/s.\\

\subsection{Visible and Infrared Adaptive Optics Imaging}\label{imaging}

\begin{figure}
\begin{center}
\includegraphics[width=\linewidth]{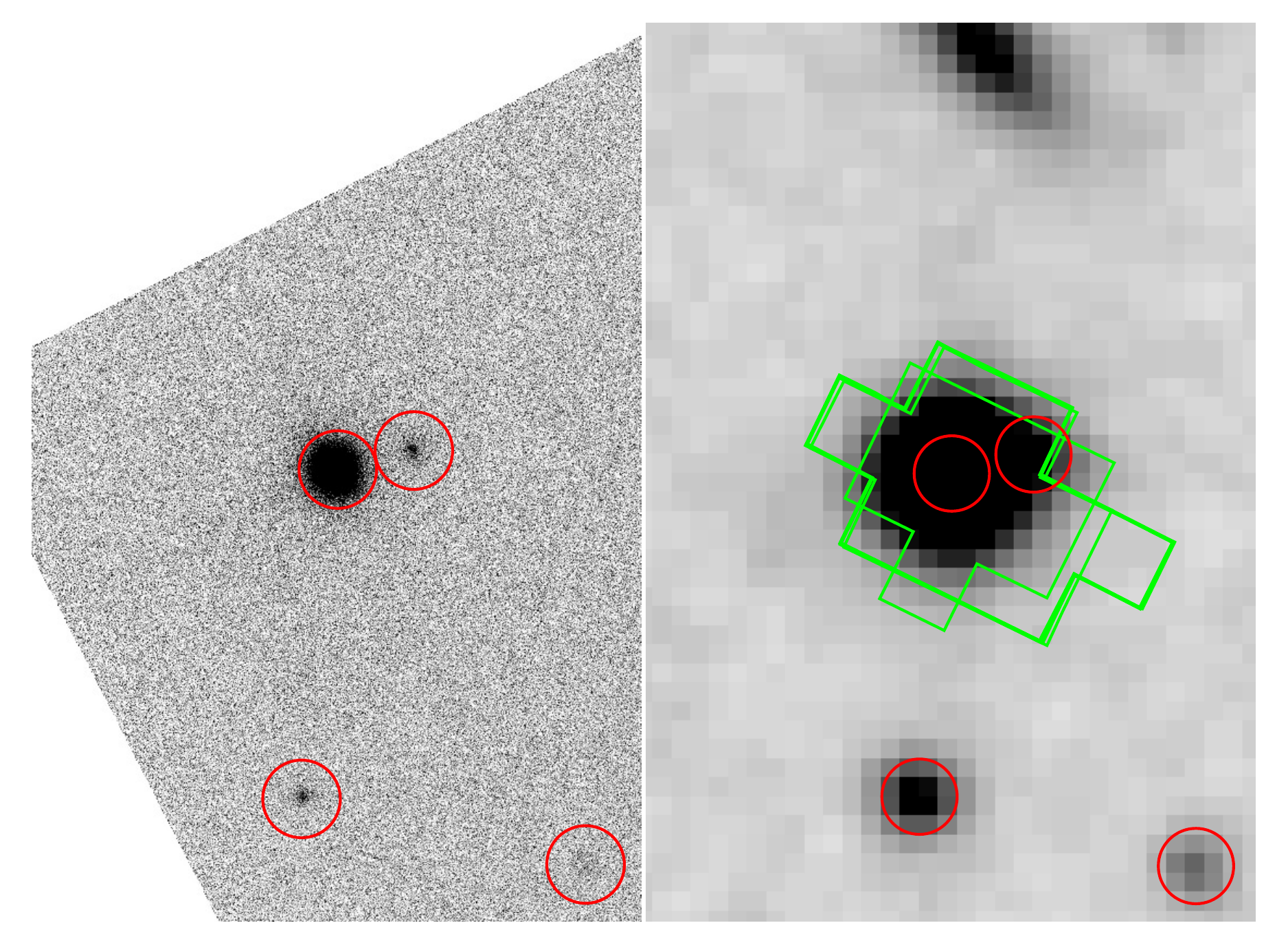}
\caption{{\it Left}: Visible-light adaptive optics image of KIC 10935310 (center object) {\it Right}: Archival RG610 (roughly $r$ band) image from the Palomar Observatory Sky Survey \citep[][]{Reid1991} including the ``postage stamps'' from the 1st four quarters (Q0, Q1, Q2 and Q3) of {\it Kepler} observations outlining the apertures used for measuring the flux from KIC 10935310 ({\it green outlines}).  North is up, east is left, and the {\it red circles} are 4" in diameter and centered on the nearby stars.}
\end{center}
\label{roboAO}
\end{figure}

As discovered by \cite{Cakirli2013a}, a faint and slightly redder object appears blended with T-Cyg1-12664 in seeing-limited images.  To determine the role of this object in the {\it Kepler} light curve and corresponding EB parameters, we acquired visible-light and infrared adaptive optics (AO) imaging of T-Cyg1-12664 using the Robo-AO system on the 60-inch Telescope at Palomar Observatory \citep{Baranec2013, Baranec2014, Law2014}.  We observed T-Cyg1-12664 on UT 2014 June 17 using a clear anti-reflective coated filter. The camera response function is spectrally limited by the E2V CCD201-20 detector response, with a steep drop off short-ward of 400 nm and long-ward of 950 nm.  This closely matches the response of the {\it Kepler} camera, which employs no filters and is primarily dictated by the CCD response.  The individual images were combined using post-facto shift-and-add processing using T-Cyg1-12664 as the tip-tilt star.  We detected the faint object at a separation of of $4\farcs12 \pm 0\farcs03$  and a position angle of $283\pm2$ degrees with respect to T-Cyg1-12664 (see Figure \ref{roboAO}).  We measured a contrast of 3.91 $\pm$ 0.10 magnitudes in the {\it Kepler} band ($K_P$) between the EB and the third object. 

On UT 2014 Sepember 3, we observed T-Cyg1-12664 in $H$ band with Robo-AO, using an engineering grade Selex ES Infrared SAPHIRA detector \citep{Finger2014} in a GL Scientific cryostat mounted to the Robo-AO near-infrared camera port \citep{Atkinson2016} with active infrared tip-tilt guiding using T-Cyg1-12664 as the reference star \citep[][]{Baranec2015}. The contrast in $H$ band was measured to be 2.74 $\pm$ 0.10 magnitudes. 

Archival photometry of T-Cyg1-12664 from the {\it Kepler} Input Catalog \citep[KIC,][]{Batalha2010} and 2MASS \citep{Cutri2003} list magnitudes of $K_P$=13.100 $\pm$ 0.03 and $H$=11.582 $\pm$ 0.015 for the blended objects.  Combining these measurements with the Robo-AO contrast measurements, we calculated magnitudes of $K_P$ = 13.129 $\pm$ 0.031 and $H$ = 11.666 $\pm$ 0.017 for the EB and $K_P$ = 17.04 $\pm$ 0.10 and $H$ = 14.41 $\pm$ 0.09 for the third object.  The color ($K_P-H=2.63 \pm 0.14$) of the third object is consistent with an early M dwarf star or a distant, intrinsically bright, and reddened evolved star.  Calculating a photometric parallax, we find that if the third object were a dwarf, it would reside roughly 150 pc more distant than the EB, though photometric parallaxes are highly uncertain.   We note that widely-separated, physically associated stars are common near EBs and are consistent with proposed scenarios for the formation of close  binaries via Kozai cycles \citep[][]{Fabrycky2007,Tokovinin2017}.

\section{Analysis and Results}\label{model}

\subsection{Light Curve Model and Fit}\label{LCmodel}

\begin{figure}
\centering
\includegraphics[width=\linewidth]{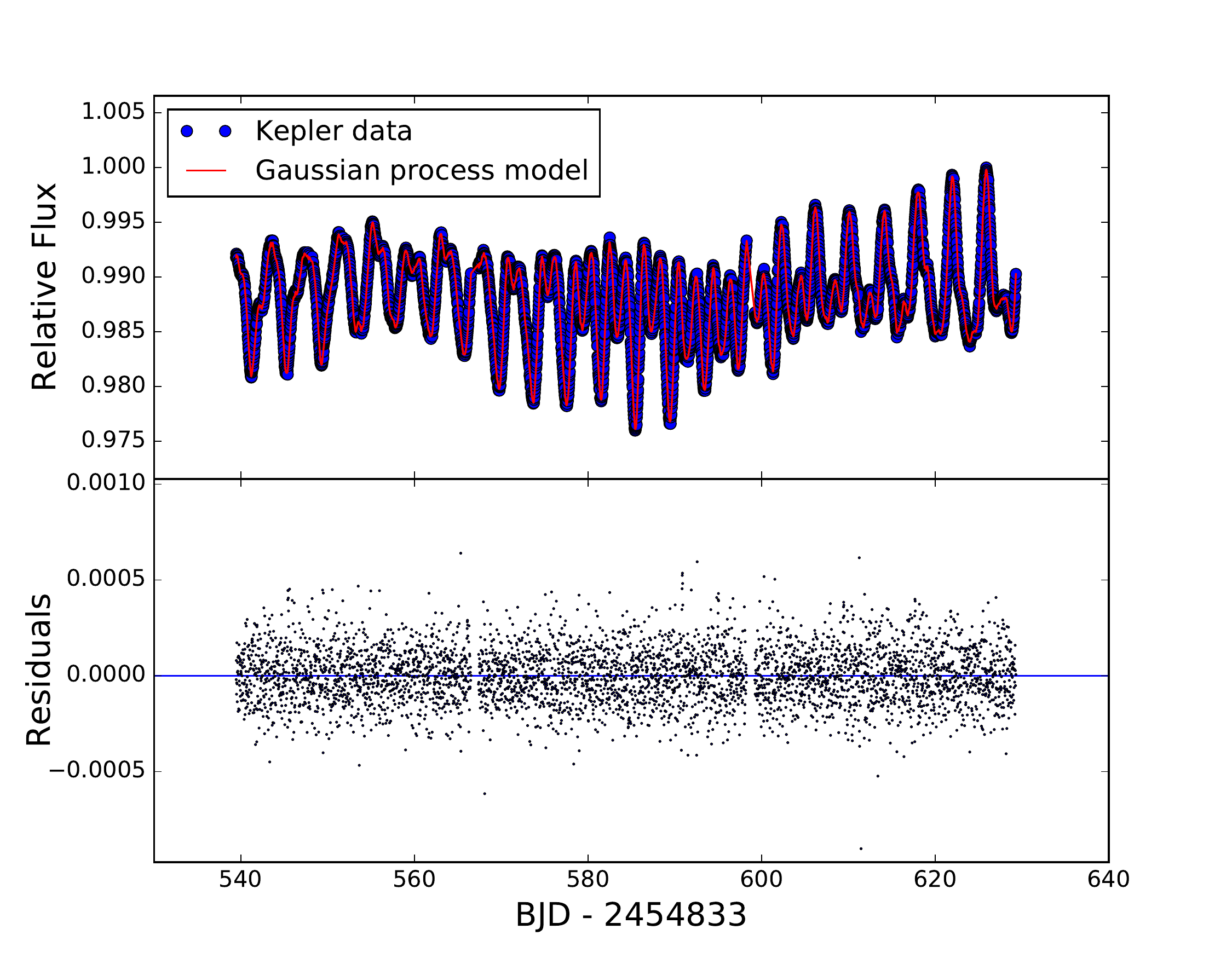}
\caption{Best-fit and residuals of the out-of-eclipse modulation in the {\it Kepler} Q6 data from \texttt{george}.}
\label{gp}
\end{figure}

\indent To study the eclipses in detail, first we modeled the out-of-eclipse modulations in order to remove their effects from the eclipse events. We discuss the causes of these modulations more in more detail in Section \ref{Discussion}. We used \texttt{george}, a Gaussian processes module written in Python \citep[][]{george}. Gaussian processes are a generalization of the normal (Gaussian) probability distribution.  The technique assumes that every point in the time series is associated with a normally distributed random variable and covariance between datapoints is constant over the dataset. The \texttt{george} software package employs `kernels' to measure the covariance between data points in the time series.
The uncertainty is calculated by taking the determinant of the $n$ x $n$ covariance matrix where $n$ is the number of data points in the time series. \\
\indent The out-of-eclipse modulations evident in the {\it Kepler} light curve show quasi-periodic behavior, that is within 3\% of the orbital period of the system.  We used the exponential-squared and the exponential-sine-squared kernels in \texttt{george} to describe the following behaviors observed in the light curve modulations: amplitude, decay/growth, and the period. We obtained the model light curve for the out-of-eclipse modulation by combining the two kernels through multiplication and fit it to the {\it Kepler} data using a Levenberg-Marquardt algorithm implemented in Python \citep[\texttt{mpfit}][]{Markwardt2009}. Figure \ref{gp} shows the out-of-eclipse light curve of the quarter 6 data and the best-fit detrending model obtained from \texttt{george}, and the resulting residuals. After detrending, we normalized the flux by dividing by the median value.

To model the eclipse events, we used the publicly available eclipsing binary modeling code \texttt{eb} written for detached eclipsing binaries by \cite{Irwin2011}. The \texttt{eb} software package creates model eclipse light curves based on 37 parameters, each of which are described in \cite{Irwin2011}. For a given set of parameters and time stamps, \texttt{eb} generates a synthetic light curve and a synthetic radial velocity curve. We chose to fit for 16 parameters and fixed the remaining 21 parameters, which describe star spots, gravity darkening, and reflection effects. The 16 free parameters that were fitted are listed in Table \ref{modelparm}. We smoothed the \texttt{eb} light curve model to account for the {\it Kepler} long-cadence integration time. \citet[][]{Coughlin2011} investigated the effect of {\it Kepler} long-cadence integration time in light curves of eclipsing binaries and found that the shape of light curve model can be changed significantly if the long-cadence integration time is not accounted for, which will result erroneous measurements in the stellar parameters. However, this effect is only shown in the systems with small relative radii sum ($<$ 0.1) and short orbital periods ($<$ 2.5 days) and we note that this is not the case of T-Cyg1-12664.\\
\indent After we detrended, normalized, and phase-folded the Kepler light curves, we employed the Levenberg-Marquardt technique and performed chi-square minimization, which was done by Python's external package, \texttt{mpfit} \citep[][]{Markwardt2009}. We ran the \texttt{mpfit} algorithm three times in order to determine the best fit values: once just varying the period and the epoch of the primary eclipse, once varying the rest of the parameters described in Table \ref{modelparm} except for the four limb-darkening parameters, and then finally varying all 16 parameters. Except for the first run, we performed each \texttt{mpfit} run using the best-fit parameters obtained from the previous run. For the first two \texttt{mpfit} runs, we did not fit the limb-darkening parameters, as limb-darkening is a higher order effect on the shape of the light curve. Instead, we adopted the square-root limb darkening law, as demonstrated in \citet[][]{Claret1998} to be superior for low mass stars like M dwarf. We set u1 = 0.63 and u2 = 0.6043 for the primary and u1 = 0.4580 and u2 = 0.6508 for the secondary, accounting the effective temperatures \citep[][]{Claret2012} based on the spectral types of \citet{Cakirli2013a}. After the second \texttt{mpfit} run, we determined that the best-fit was close enough to treat the limb-darkening coefficients as free parameters. For the \texttt{mpfit} to return a good-fit, we had to be careful with choosing the step sizes. 
We excluded the majority of the out-of-eclipse light curve as they were the dominant noise source in the $\chi^2$ calculation and as the flattened out-of-eclipse fluxes had no information on the component stars. \\
\indent To further refine the fit and determine reliable uncertainties for the individual parameters, we employed a Markov chain Monte Carlo (MCMC) algorithm.  We used Python's external MCMC package, \texttt{emcee}, written by  \citet[][]{Foreman-Mackey2013}.  We used the best-fit parameters from the last \texttt{mpfit} fit as the starting parameters in the MCMC chains.  We employed 100 chains, each with 10000 steps, and assumed uniform priors on all parameters. We varied $T_0$, $P$, $J$, $\cos{i}$, $e \cos{\omega}$, $e \sin{\omega}$, $(R1+R2)/a$, $R2/R1$ and four limb-darkening parameters, two for each component. For the limb-darkening parameters, we stepped in the $q_1$ and $q_2$ parametrization of limb-darkening, developed by \cite{Kipping2013}, rather than the linear and quadratic coefficients.  The $q_1$ and $q_2$ parametrization of limb-darkening forces all possible combinations of the parameters to be physical, as long as both values are between 0 and 1. For the third light, $L3$, we set it to the value we directly measured from the AO imaging.

Although the \texttt{eb} model takes $e\cos{\omega}$ and $e\sin{\omega}$ as free parameters, we stepped in $\sqrt{e} \cos{\omega}$ and $\sqrt{e} \sin{\omega}$ as \cite{Eastman2013} argue this is more efficient.
Once completed, we discarded the ``burn-in'' and took the maximum likelihood parameters as the best-fit values, and the standard deviations of the parameter distributions as the uncertainties.  We report our best fit values and uncertainties in Table \ref{fitted_parameters}. \\

\subsection{Radial Velocity Model and Fit}

To independently measure the masses of each component, we did not combine our radial velocity data with the previously published radial velocity data. Moreover, instead of fitting the light curve and the radial velocity data simultaneously, we chose to fit the photometric and the spectroscopic data individually.  We found that simultaneous fitting resulted in poor fits to the radial velocity data, because the number of data points in the {\it Kepler} data far outweigh the radial velocity data.  

\indent As we briefly mentioned in Section \ref{LCmodel}, the \texttt{eb} software package outputs an RV model, which we can use when performing the radial velocity fit of the 16 free parameters in Table \ref{modelparm}.  Parameters that affect the RV model are the orbital period (P), the epoch of the primary mid-eclipse ($t_0$), the mass ratio ($q$), $K_{tot}/c$, the systematic velocity ($\gamma$), $e\cos{\omega}$, and $e\sin{\omega}$. The orbital period ($P$) and the epoch of the primary mid-eclipse ($t_0$) were fixed to the values from the best-fit values from the light curve. Since the light curve has more data points than the radial velocity data and samples in a finer step, for $e \cos{\omega}$ and $e \sin{\omega}$, we took their posterior distribution from the light curve fit as the priors for the MCMC run in the radial velocity fitting. 

Unlike in the light curve fitting, we did not employ \texttt{mpfit}, since we had good starting points for all the parameters from the light curve fit. In order to confirm the validity of our code, we tried fitting the \cite{Devor2008}, \cite{Cakirli2013a}, and the \cite{Iglesias2017} radial velocity points, which is the same set of radial velocity data that \cite{Iglesias2017} had in their fit. Our calculated mass ratio, individual mass, the radial velocity semi-amplitudes, and the sum of the semi-amplitudes match well with those of \cite{Iglesias2017} within the uncertainties. Following this validity check, we fitted IGRINS H-band, NIRSPEC K-band, and HIRES radial velocity points alone, ignoring the measurements of \citet{Cakirli2013a} or \citet{Iglesias2017}.

\subsection{Results}
\begin{figure}
\centering
\includegraphics[width=\linewidth]{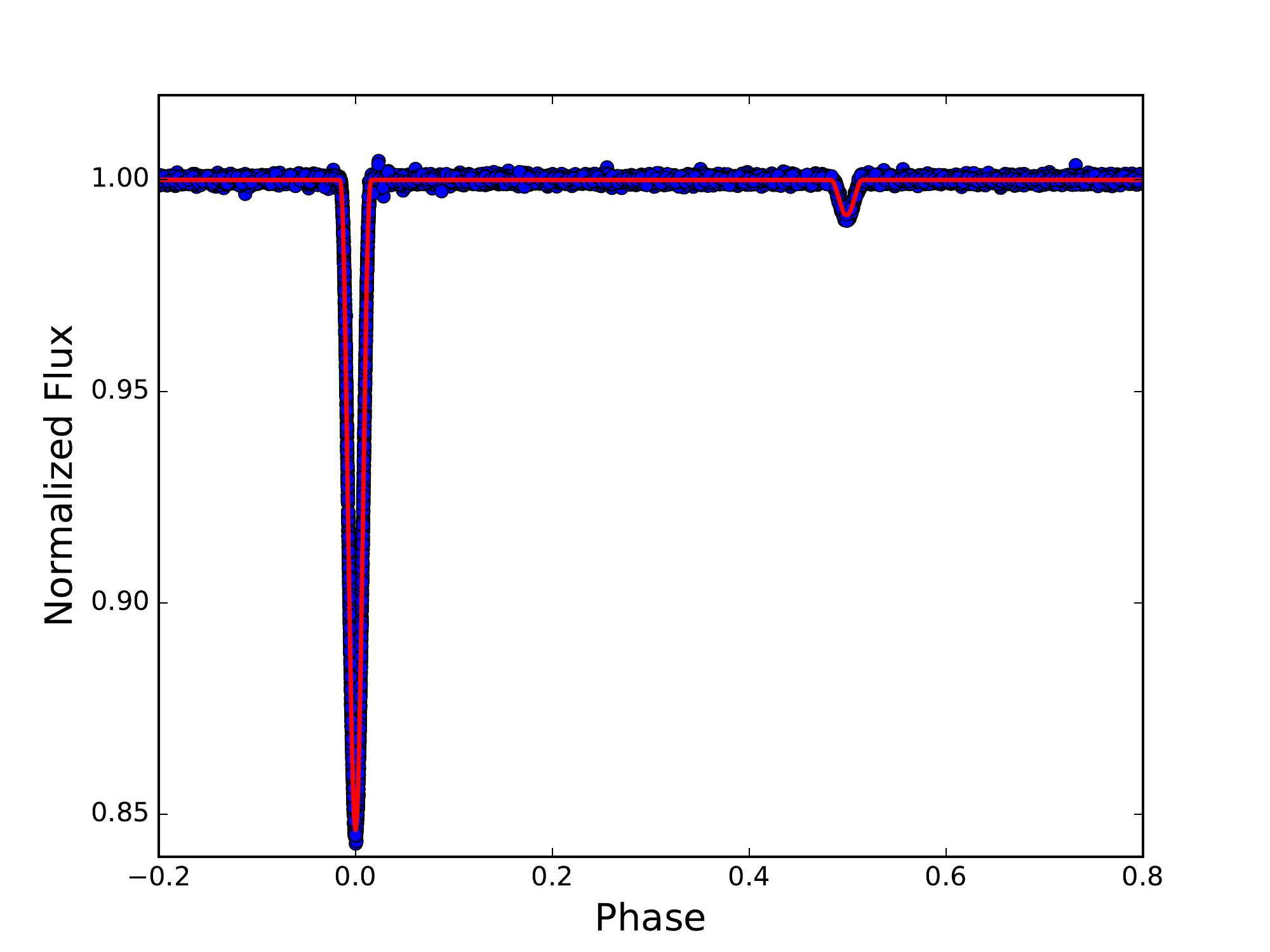}
\caption{Detrended and phase folded {\it Kepler} light curves for all quarters except for 7, 11, 12 and 15 with the best-fit model (red line).}
\label{model_fit_full}
\end{figure}

\begin{figure*}
\centering
\includegraphics[width=7.0in]{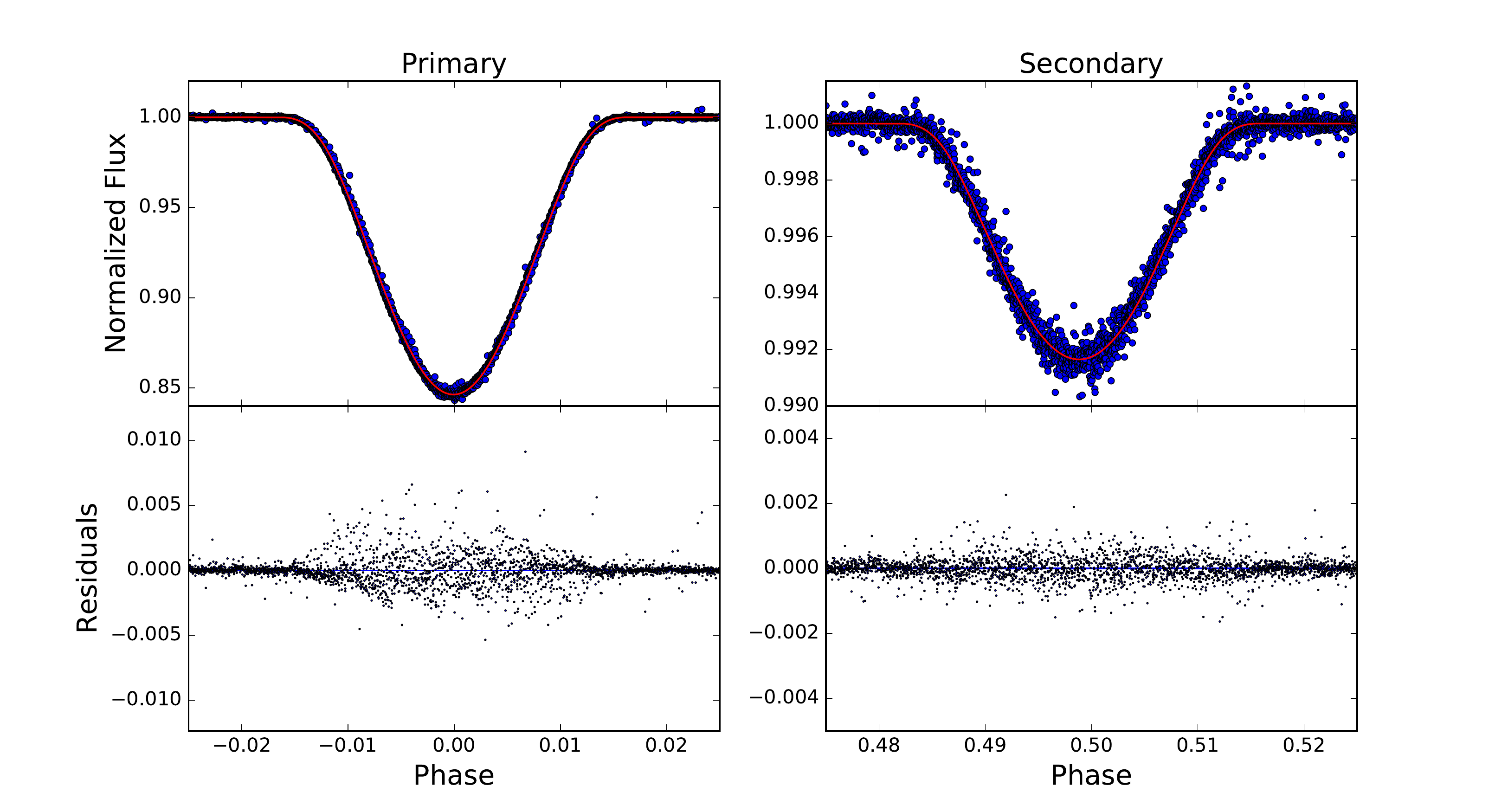}
\caption{Zoom-in of the primary and the secondary eclipses. The top panels show detrended and phase folded {\it Kepler} data with their best fit and the bottom panels show the residuals.}
\label{model_fit}
\end{figure*}

\begin{figure*}
\centering
\includegraphics[width=0.75\linewidth]{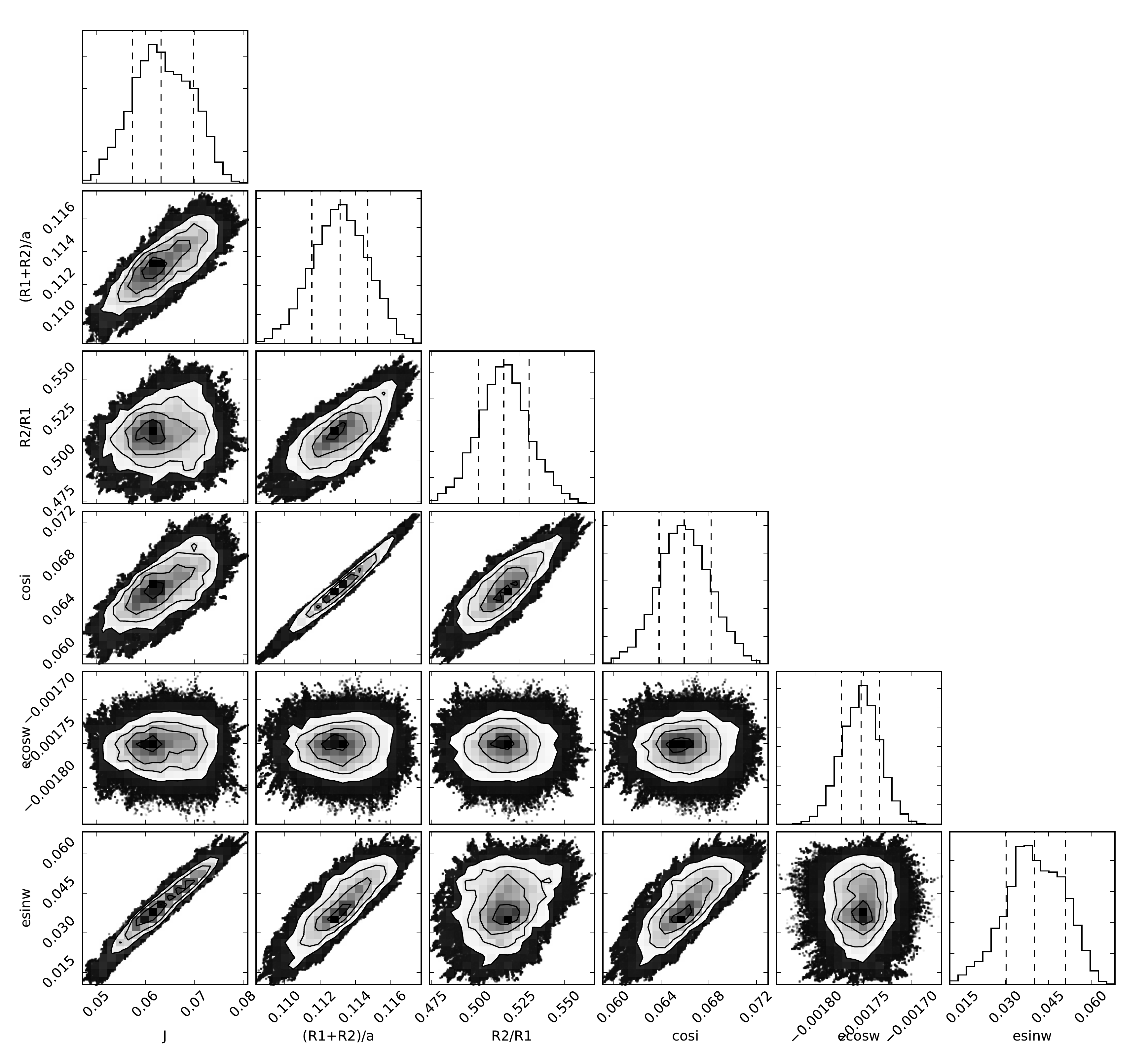}
\caption{Triangle plot of the light curve fit. The histogram and the contour plots show density of MCMC iterations. The dashed lines in the histogram mark 16$^{th}$, 50$^{th}$, and 84$^{th}$ percentiles of the samples in the marginalized distributions. 
See Table \ref{modelparm} for descriptions of the fitted parameters.}
\label{LC_corner}
\end{figure*}

\begin{figure*}
\centering
\includegraphics[width=0.75\linewidth]{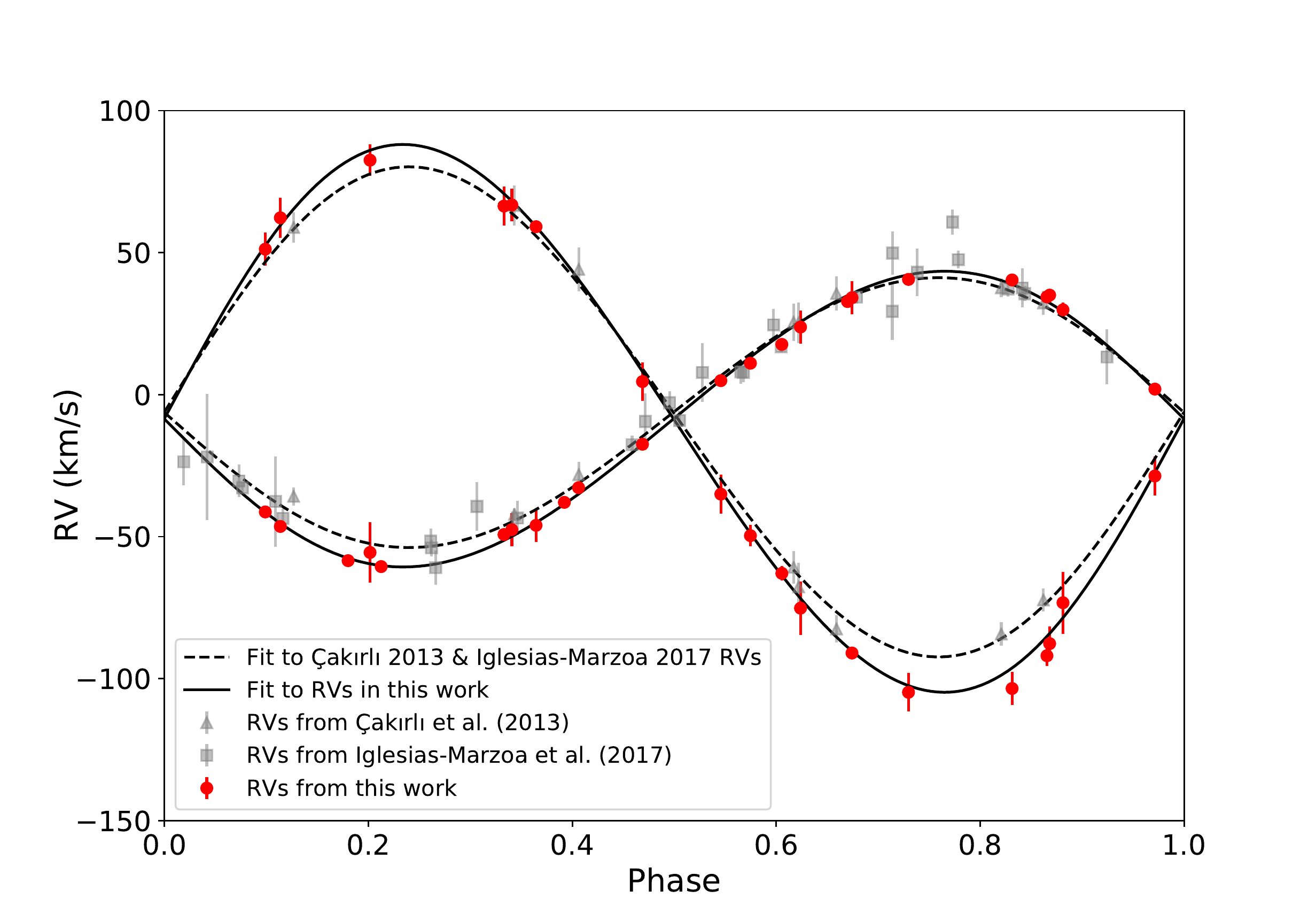}
\caption{Best-fit to the radial velocity data. The solid line is best-fit with IGRINS, NIRSPEC, and HIRES data and the dashed line is the best-fit with the \cite{Cakirli2013a} and \cite{Iglesias2017} data. The calculated radial velocity semi-amplitudes are $K_1 = 52.3 \pm 1.2 $km/s for the primary and $K_2 = 97.3 \pm 1.3$km/s for the secondary using data from this work.} 
\label{rvfit}
\end{figure*}

Figure \ref{model_fit_full}, Figure \ref{model_fit}, and Figure \ref{rvfit} show the best fit models we obtained using \texttt{eb}. 
Figure \ref{model_fit_full} shows the phase-folded {\it Kepler} light curve data with the best-fit model plotted in red. Figure \ref{model_fit} shows the zoomed-in region around each eclipse with the best-fit (top panel) and the residuals (bottom panel). We report the parameters from the model with the highest likelihood as our best fit values. We adopted the standard deviation of the MCMC chains as the uncertainty for all of the parameters with symmetric posterior distribution. However, for esinw, the distributions are not symmetric and we choose to use the 34.1 percentile around the highest likelihood value.
 Figure \ref{LC_corner} shows the triangle plot from the light curve MCMC run. Figure \ref{rvfit} shows the primary (in blue) and the secondary (in green) data points from all available radial velocity data with the best-fit model in red. \\
\indent Table \ref{fitted_parameters} shows the fitted and the calculated parameters from the light curve and radial velocity fitting. We measured $K_1 = 52.3 \pm 1.2 $km/s and $K_2 = 97.3 \pm 1.3$ km/s. For the masses and the radii, we measured $M_1 = 0.92 \pm 0.05M_{\sun}$ and $R_1 = 0.92 \pm 0.03 R_{\sun}$ for the primary and $M_2 = 0.50 \pm 0.03 M_{\sun}$ and $R_2 = 0.47 \pm 0.04 R_{\sun}$ for the secondary. Our measured masses and radii are different from the previous two groups' results, which we discuss in the following section.\\

\begin{table*}
\begin{center}
\caption{Modeling Parameters}
\begin{tabular}{@{}l c}
\hline
Parameter & Description \\
\hline
$J$ & Central surface brightness ratio \\
$(R_1 + R_2)/a$ & Fractional sum of the radii over the semi-major axis\\
$R_2/R_1$ & Radii ratio\\
$\cos{i}$ & Cosine of orbital inclination \\
$P$ (days) & Orbital period in days \\
$T_0$ (BJD) & Primary mid-eclipse \\
$e \cos{\omega}$ & Orbital eccentricity $\times$ cosine of argument of periastron  \\
$e \sin{\omega}$ & Orbital eccentricity $\times$ sine of argument of periastron\\
$L3$ & Third light contribution from a nearby companion \\
$\gamma$ (km s$^{-1}$) & Center of mass velocity of the system \\
$q$ & Mass ratio ($M_2$/$M_1$) \\
K$_{tot}$ & Sum of the radial velocity semi-amplitude\\
u$_{K_p}$ & Linear limb-darkening coefficient in {\it Kepler} band\\
u'$_{K_p}$ & Square root limb-darkening coefficient in {\it Kepler} band\\
\hline
\label{modelparm}
\end{tabular}
\end{center}
\end{table*}

\begin{table*}
\begin{center}
\caption{Measured radial velocities for the primary and the secondary stars}
\begin{tabular}{@{}l c c c c c}
\hline
BJD & $V_1$ (km/s) & $\sigma_1$ (km/s) & $V_2$ (km/s) & $\sigma_2$ (km/s) & Instrument\\
\hline
2457678.606272 & -43.4 & 1.0 & 49.1 & 5.8 & IGRINS\\
2457678.667269 & -48.5 & 0.5 & 60.2 & 7.0 & IGRINS\\
2457679.604463 & -50.4 & 5.9 & 63.9 & 5.8 & IGRINS\\
2457679.702549 & -48.1 & 0.5 & 57.0 & 0.4 & IGRINS\\ 
2457680.570521 & 9.0 & 1.6 & -51.7 & 3.8 & IGRINS\\
2457680.697738 & 17.1 & 1.1 & -63.5 & 5.8 & IGRINS\\
2457680.773412 & 21.7 & 5.9 & -77.3 & 9.4 & IGRINS\\
2454311.884614 & 34.1 & 5.9 & -91.0 & 1.9 & NIRSPEC\\
2456845.013664 & -55.6 & 10.7 & 82.6 & 5.5 & NIRSPEC\\
2456851.948214 & 29.9 & 2.6 & -73.3 & 10.9 & NIRSPEC\\
2456827.110291 & 34.4 & 0.4 & -92.0 & 3.7 & HIRES\\
2456829.040765 & 49.3 & 1.1 & 66.4 & 6.8 & HIRES\\
2456829.918712 & 4.9 & 0.8 & -35.1 & 6.8 & HIRES\\
2456831.098019 & 40.4 & 0.8 & -103.5 & 5.9 & HIRES\\
2456843.064820 & 40.6 & 0.5 & -104.8 & 6.8 & HIRES\\ 
2456844.062487 & 1.9 & 1.7 & -28.7 & 6.8 & HIRES\\
2456845.058524 & -60.6 & 0.7 & - & - & HIRES\\
2456846.116576 & -17.5 & 0.6 & 4.6 & 6.8 & HIRES\\
2456846.946306 & 32.8 & 1.14 & - & - & HIRES\\
2456849.053336 & -58.5 & 0.6 & - & - & HIRES\\ 
2456849.986511 & -32.8 & 0.6 & - & - & HIRES\\
2456851.894018 &  35.0 & 0.8 & -87.7 & 6.1 & HIRES\\
2456854.057728 & -37.9 & 1.5 & - & - & HIRES\\
\hline
\label{rv}
\end{tabular}
\end{center}
\end{table*}

\begin{figure*}
\centering
\includegraphics[width=0.6\linewidth]{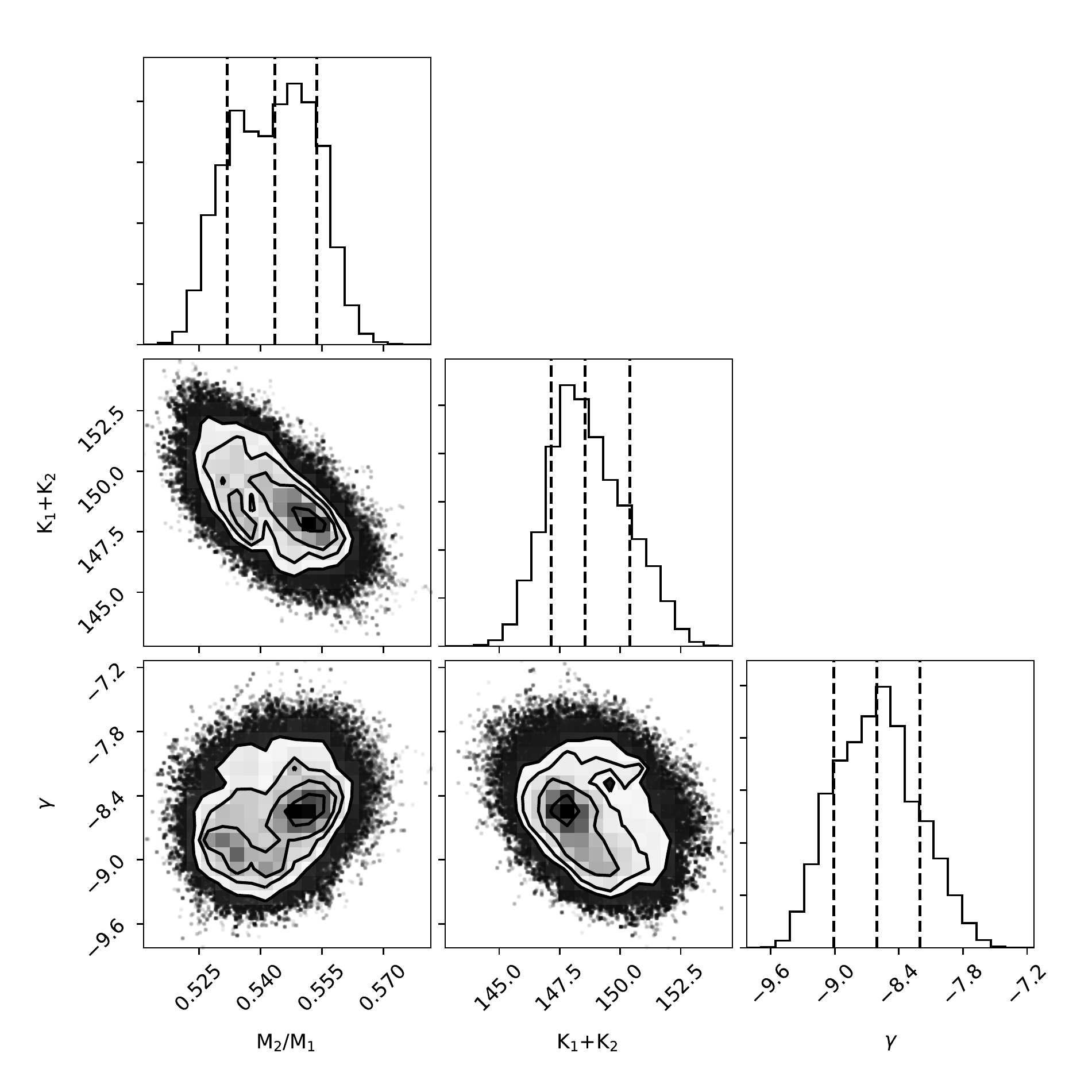}
\caption{Triangle plot showing the density of MCMC iterations for the radial velocity fit. See Figure \ref{LC_corner} for the descriptions. }
\label{RV_corner}
\end{figure*}

\begin{table*}
\begin{center}
\caption{Parameters for T-Cyg1-12664 (this work).}
\begin{tabular}{@{}l c c}
\hline
Fitted & Primary & Secondary \\
\hline
$J$ & \multicolumn{2}{c}{0.0675 $\pm$ 0.0069} \\
$(R_1 + R_2)/a$ & \multicolumn{2}{c}{0.1138 $\pm$ 0.0023} \\
$R_2/R_1$ & \multicolumn{2}{c}{0.5161 $\pm$ 0.0224} \\
$cosi$ & \multicolumn{2}{c}{0.0666 $\pm$ 0.0034} \\
$P$ (days) & \multicolumn{2}{c}{4.12879671 $\pm$ 0.00000003} \\
$T_0$ (BJD) & \multicolumn{2}{c}{2454957.32116092 $\pm$ 0.0000051} \\
$e \cos{\omega}$ & \multicolumn{2}{c}{-0.00176 $\pm$ 0.00002} \\
$e \sin{\omega}$ & \multicolumn{2}{c}{$0.0438^{+0.0121}_{-0.0089} $} \\
$L3$ & \multicolumn{2}{c}{0.0265 $\pm$ 0.0025 (fixed)}\\
$\gamma$ (km s$^{-1}$) & \multicolumn{2}{c}{-8.6 $\pm$ 0.4} \\
q & \multicolumn{2}{c}{0.54 $\pm $ 0.01}\\
$K_{tot}$ & \multicolumn{2}{c}{149.6 $\pm$ 1.6}\\
$u_{K_P}$ &$ 0.053^{+0.039}_{-0.030}$ & $0.757^{+0.242}_{-0.171}$ \\
$u'_{K_P}$ & $0.944^{+0.033}_{-0.057} $ & $0.025^{+0.107}_{-0.025}$\\
\hline
Calculated & Primary & Secondary \\
\hline
$e$ & \multicolumn{2}{c}{$0.0439^{+0.0024}_{-0.0026}$}\\
$i$ ($^\circ$) & \multicolumn{2}{c}{86.20 $\pm$ 0.20}  \\
$a_{\rm tot}$ (R$_{\sun}$) & \multicolumn{2}{c}{12.23 $\pm$ 0.16} \\
$K$ (km s$^{-1}$) & 52.3 $\pm$ 1.2 & 97.3 $\pm$ 1.3 \\
$M$ ($\rm M_{\sun}$) & 0.92 $\pm$ 0.05 & 0.50 $\pm$ 0.03 \\
$R$ ($\rm R_{\sun}$)  & 0.92$\pm$ 0.03 & 0.47 $\pm$ 0.04 \\
$K_P$ (mag) & 13.141 $\pm$ 0.031 & 18.066 $\pm$ 0.031 \\
\hline
\label{fitted_parameters}
\end{tabular}
\end{center}
\end{table*}

\section{Discussion}\label{Discussion}

\subsection{Third Light}
Comparing to the stellar parameters of {\it Kepler} targets in \cite{Huber2014}, the resulting $K_P$ magnitudes for the primary and secondary are consistent with the fitted masses and radii and indicate an mid-G and mid-M dwarf EB at a distance of  $\sim$460 pc, significantly further than the estimated distance of \cite{Cakirli2013a}.  The color for the third object is consistent with an early M dwarf at a distance of $\sim$610 pc, or a distant, evolved and reddened star. In either case the third object would not be associated with the EB; however, the uncertainty in distance is large as it is based entirely on a single color that is expected to be degenerate with stellar metallicity. There remains a distinct possibility that the third object is an associated early M dwarf star. The third light contribution we measured is different by $\sim$1.4 sigma in comparison with what \cite{Iglesias2017} reported. We attribute this difference to our inclusion of adaptive optics imaging, which provides a direct measurement of the third light, rather than fitting it as a model parameter in the light curve.

\subsection{Out-of-eclipse Modulations}
In the {\it Kepler} light curve, the causes of modulations in out-of-eclipse data can be star spots, reflection effects, ellipsoidal variations, beaming effects, and gravity-darkening. The reflection, ellipsoidal, and beaming signals for our best fit parameters are at least one order of magnitude less than the observed modulations \citep[][]{rep}. Gravity darkening, according to von Zeipel Theorem \citep[][]{vonZeipel1924}, is significant in stars that are hot enough to have radiative envelopes (earlier than F type), which is not the case for T-Cyg1-12664. Therefore, the dominant cause of out-of-eclipse modulation must be star spots.

\subsection{Eccentricity and Age}
The eccentricity of the orbit is non-zero. The effect of non-zero eccentricity is shown in the midtime of the secondary eclipse, which slightly departs from 0.5 in orbital phase. Eccentricity and the argument of periastron ($\omega$) determine the time interval between the primary and the secondary eclipse (ecos$\omega$) and the duration of eclipse (esin$\omega$).
For stars with a convective envelope, the circularization timescale ($\tau_{circ}$) and the synchronization ($\tau_{sync}$) timescale are proportional to $(a/R_1)^8$ and $(a/R_1)^6$, respectively, where a is the binary semi-major axis and $R_1$ is the radius of the primary component \citep[][]{Zahn1975}. For T-Cyg1-12664, these timescales are $\tau_{sync}$ $\simeq$ 5.8 Myr and $\tau_{circ}$ $\simeq$ 1.1 Gyr. As evident from the out-of-eclipse modulations in the {\it Kepler} light curve, the binary orbit is nearly synchronized. However, the binary orbit is not circularized as the eccentricity of the orbit is non-zero.

\subsection{Comparison with the previous publications}
Our reported masses and radii differ from the previous two publications. We attribute the discrepancies to the difference in the radial velocity measurements. Our spectroscopic data completely cover the orbital period and allow the radial velocity measurements of both components. In fact, \cite{Iglesias2017} stated that future radial velocity observations, specifically near-infrared observations, would likely further refine the system parameters.

T-Cyg1-12664 is a spotted system. Starspots on a rotating photosphere can introduce radial velocity variations \citep[e.g.][]{Andersen2015}.  \cite{Gagne2016} investigated the effect of star spots on radial velocity measurements and found that an active star shows a long-term radial velocity variations of 25-50 m/s in the near-infrared. In the near-infrared, spot-induced radial velocity signal is significantly reduced thanks to the lower contrast between spots on the photosphere at longer wavelengths. \citet[][]{Reiners2010} showed that spot-induced radial velocity variations have a $\lambda^{-1}$ dependence, where $\lambda$ is the observed wavelength. Both combined, the radial velocity signal from spots in our measurements are not significant and well within the measurement uncertainties, even for the visible-wavelength HIRES observations.

The primary and the secondary components are different in spectral type.  A mismatch in radial velocity templates can cause offsets in the radial velocity zero point. For the IGRINS and NIRSPEC spectra, we used a combination of a G5 and a M3 template, and for the HIRES spectra, we used a combination of a K1 and M3.5 template. The data sets have a consistent radial velocity zero-points despite using different templates. For this reason, we do not believe this is responsible for the discrepancy with the two previous studies.

\cite{Cakirli2013a} reports the orbital inclination angle of $i = 83.84^\circ \pm 0.04^\circ$, corresponding to grazing eclipses. For grazing eclipses, the extracted radius ratio from the photometry alone is not well constrained and is degenerate with other parameters. To better constrain the radius ratio, spectroscopic light ratios must be provided, which was not the case in \cite{Cakirli2013a}. From our analysis, we measured $i = 86.20^\circ \pm 0.20^\circ$, which is nearly an edge-on orbital configuration, and our result is much less affected by the degenerate radius ratio. However, we were not able to measure the spectroscopic light ratio from our data and this remains as a caveat to our measurement.
The measured masses and radii for both components indicate that neither of the stars is inflated and the values agree well with the predictions of stellar evolutionary models.  \cite{Iglesias2017} reported the spectral type of the primary to be G6 main-sequence star. However, their reported primary mass is 0.680$M_{\sun}$, which is low for a typical a main-sequence G6 type star. Our measurement for the primary component mass indicates the primary star is a solar type star. For the secondary component, our measurement corresponds to an early-M dwarf star.

\subsection{A Note on Effective Temperature}
We note that unlike other eclipsing-binary fitting procedures, our method did not fit for the effective temperatures of the component stars, which would result in measured spectral types and a semi-empirical distance to the system.  We purposefully do not fit for effective temperature, instead fitting for the central surface brightness ratio between the two stars in the {\it Kepler} band.  We see this as an advantage.  To determine the effective temperatures of the component stars, we would have to invoke bolometric corrections that depend on accurate atmospheric models of the stars.  Atmospheric models of low-mass stars are known to disagree with spectroscopic observations due to the many molecular opacities required \citep[][]{Allard2012}. A recent investigation by \citet{Veyette2016} showed that the carbon-to-oxygen ratio of a low-mass star can dramatically change model spectra of low-mass stars, even at fixed effective temperature, which would also affect the reported effective temperature. Instead, by reporting specifically the ratio of component stars' central surface brightnesses in a well defined band, we remove the assumptions about metallicity, carbon-to-oxygen ratio and particular molecular opacity tables.

\section{Conclusions}
\begin{figure*}
\centering
\includegraphics[width=0.75\linewidth]{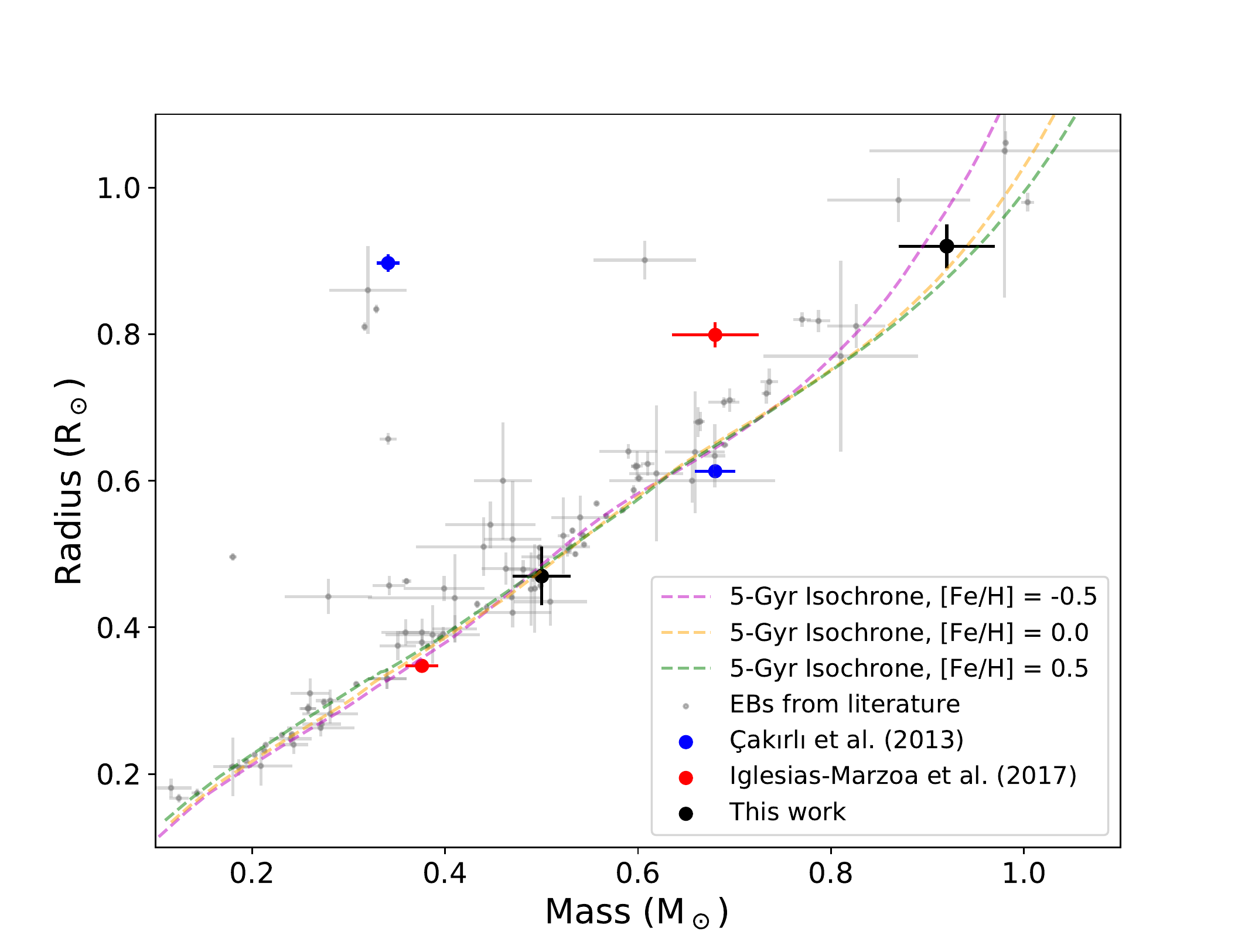}
\caption{Mass vs. radius for main-sequence low-mass EBs ({\it gray circles}, see \S \ref{intro} for references) with the former values for T-Cyg1-12664 shown as {\it large red} and {\it large blue circles} and the revised values as {\it large black circles}.  We include predictions for 5-Gyr-old stars from the Dartmouth evolutionary isochrones as dashed lines for three metallicities: [M/H] = -0.5 ({\it magenta}), 0.0 ({\it orange}) and +0.5 ({\it green}) \citep{Dotter2008}.}
\label{mass_rad_figure}
\end{figure*}

Figure \ref{mass_rad_figure} plots mass versus radius for published low-mass stars in EBs and the former and revised determinations for the components of T-Cyg1-12664.  We have revised the mass and the radius of both components by a substantial amount. For the primary star, we revised the mass from  0.680 $\pm$ 0.045 $M_{\sun}$ to  0.92 $\pm$ 0.05 $M_{\Sun}$ and the radius from 0.799 $\pm$ 0.017 $R_{\Sun}$ to 0.92 $\pm$ 0.03 $R_{\Sun}$. For the secondary star, we revised the mass from 0.376 $\pm$ 0.017 $M_{\sun}$ to  0.50 $\pm$ 0.03 $M_{\Sun}$ and the radius from 0.35 $\pm$ 0.01 $R_{\Sun}$ to 0.47 $\pm$ 0.04 $R_{\Sun}$. The measured masses and radii indicate that neither stars are inflated and the values agree well with the predictions of stellar evolutionary models.\\
\indent It is not entirely clear why the radii from this work and \cite{Cakirli2013a} are so discrepant despite using nearly identical data.  However, \cite{Cakirli2013a} did not mention fitting an eccentricity, which is clearly non-zero by inspection of the secondary mid-eclipse time. \cite{Cakirli2013a} also mention fitting {\it Kepler} short-cadence observations of the target. We were unable to find any short-cadence observations of the target in MAST. Given the spot crossing events during primary eclipse, and that the secondary component contributes only $\sim$1\% of the flux in the {\it Kepler} light curve, but that the light curve shows $\sim$2\% rotational spot modulation, we conclude that the primary star is in fact highly magnetically active.

\clearpage
\section*{Acknowledgments}

The authors wish to thank Andrew Howard and Howard Isaacson for incorporating the HIRES observations into the California Planet Search program.  E.H thanks Paul Dalba, Mark Veyette, and Alvaro Ribas for their advice on MCMC fitting and Aurora Kesseli for her help with the IGRINS observing runs.
E.H, P.S.M, and J.J.S acknowledge support from the NASA Exoplanet Research Program (XRP) under Grant No. NNX15AG08G issued through the Science Mission Directorate. C.B. acknowledges support from the Alfred P. Sloan Foundation. D.A. is supported by a NASA Space Technology Research Fellowship Grant No. NNX13AL75H.  

This research involved use of the Discovery Channel Telescope at Lowell Observatory, supported by Discovery Communications, Inc., Boston University, the University of Maryland, the University of Toledo and Northern Arizona University.  This research involved use of the Immersion Grating Infrared Spectrometer (IGRINS) that was developed under a collaboration between the University of Texas at Austin and the Korea Astronomy and Space Science Institute (KASI) with the financial support of the US National Science Foundation under grant AST-1229522, of the University of Texas at Austin, and of the Korean GMT Project of KASI.  This paper includes data taken at The McDonald Observatory of The University of Texas at Austin.

This research made use of the Robo-AO system. The Robo-AO system is supported by collaborating partner institutions, the California Institute of Technology and the Inter-University Centre for Astronomy and Astrophysics, by the NSF under Grant Nos. AST-0906060, AST-0960343 and AST-1207891, by the Mount Cuba Astronomical Foundation, and by a gift from Samuel Oschin. Development and characterization of the SAPHIRA detectors at the University of  Hawai`i is sponsored by the National Science Foundation under Grant No. AST-1106391 and by NASA ROSES APRA award No. NNX 13AC13G. 

Some of the data presented in this paper were obtained from the Mikulski Archive for Space Telescopes (MAST). STScI is operated by the Association of Universities for Research in Astronomy, Inc., under NASA contract NAS5-26555. Support for MAST for non-HST data is provided by the NASA Office of Space Science via grant NNX09AF08G and by other grants and contracts.  This paper includes data collected by the {\it Kepler} Mission. Funding for the {\it Kepler} Mission is provided by the NASA Science Mission Directorate.  This research involved use of the Massachusetts Green High Performance Computing Center in Holyoke, MA. 

Some of the data presented herein were obtained at the W. M. Keck Observatory, which is operated as a scientific partnership among the California Institute of Technology, the University of California and the National Aeronautics and Space Administration. The Observatory was made possible by the generous financial support of the W. M. Keck Foundation.  This research has made use of the Keck Observatory Archive (KOA), which is operated by the W. M. Keck Observatory and the NASA Exoplanet Science Institute (NExScI), under contract with the National Aeronautics and Space Administration.  The authors wish to recognize and acknowledge the very significant cultural role and reverence that the summit of Maunakea has always had within the indigenous Hawaiian community.  We are most fortunate to have the opportunity to conduct observations from this mountain.

\vspace{5mm}
\facilities{DCT (IGRINS), Keck:I (HIRES), Keck:II (NIRSPEC), Kepler, PO:1.5m (Robo-AO), Smith (IGRINS)}

\software{
\texttt{eb} \citep{Irwin2011},  
\texttt{emcee} \citep[][]{Foreman-Mackey2013},
\texttt{george} \citep[][]{george},
\texttt{mpfit} \citep{Markwardt2009},
\texttt{xtellcor} \citep[][]{xtellcor}
 }

\bibliography{references}
\bibliographystyle{aasjournal}

\end{document}